# Resonantly Enhanced Electric-Field Sensing of Etchless Thin Film Lithium Niobate via Quasibound Sates in the Continuum

*Zhijin Huang[a, *], Junzhong Wang[a], Lifang Yuan[a], Kaixiang Shen[a], Qianqian Li[a], Juan Wang[a].*

[a]*School of Electronics and Communication, Guangdong Mechanical and Electrical Polytechnic, Guangzhou 510550, China.*

*Corresponding author: 2022010010@gdmec.edu.cn*

ABSTRACT

Electric field detection has been widely utilized in many fields such as scientific research and integrated circuits. To enhance the tuning sensitivity of E-field sensor, in this paper, we theoretically proposed a highly sensitive E-field sensor composed of etchless lithium niobate (LN) material and hybrid coupling-grating systems in the visible near-infrared regime. Such configuration supports high-quality factor quasi-BIC resonance, which generates strong localized field confinement. Due to the large electro-optic coefficient of LN material, one can shift the wavelength and reflection ratio of resonance by tuning the refractive index of LN material. An analytical theory is carried out to explain the relationship between the refractive index variations and the applied voltages, so we successfully obtained a tuning sensitivity of 40.8 nm/V and a minimum detectable electric field amplitude of 24.5 mV with wavelength resolution of 1 nm. Due to the low parasitic capacitance of LN material and high conductivity of gold film and ITO layer which are utilized as the electrodes, the 3dB bandwidth of the devices should exceed 154 GHz. And we believe that such a surface-normal E-field sensor has extensive potential for the extremely weak electric field detection.

## 1. Introduction

The accurate measurement of extremely electric field has an increasing demand in many applications[1–3], ranging from process control of industrial equipment, medical instrument to the EEG or ECG signal detection. Oftentimes, traditional E-field sensor is made of CMOS-compatible materials such as Si, Si₃N₄ coated with organic electro-optical materials which limited by its stability[4–8]. Therefore, it is necessary to select better EO material for developing highly sensitive E-field sensor. Among the EO materials, lithium niobate (LN) is an attractive candidate for the development of highly sensitive E-field sensor because of commercial availability and multi-functional properties, including excellent visible to mid-infrared transparency, large EO coefficients[3,9,10]. In the past decades, LN-based E-field sensor has seen many applications in different configurations of micro and nanowaveguides such as Annealed Proton Exchange (APE) LN waveguide[11], one-dimensional LN photonic crystal[12], and straight LN nanowaveguides[13,14]. These devices operate by designing traveling-wave electrodes on both sides of the waveguide, and the external electric field is loaded on the electrodes to change the refractive index of LN to modulate the waveguide mode. Such a structure can also achieve better modulation effect by increasing the size to achieve lower applied voltage. However, on the one hand, this has resulted in a larger size, which is not suitable for the development trend of compactness and miniaturization; The distance between the traveling wave electrodes on the other side is usually tens of microns, so a higher applied voltage is required to obtain an electric field strength sufficient to change the LN refractive index.

Bound states in the continuum (BIC) are a special kind of optical resonance state which has an infinitely high-quality factor[15–18], but cannot be directly triggered by plane wave. So, one can introduce the loss of optical system to transition ideal BIC into quasi-BIC resonance[19]. Although quasi-BIC is a leaky state, it still has very large quality factor and extremely strong optical field localization ability, which can greatly enhance the interaction between light and matter[4,20], and are often utilized to realize biosensor[21], refractive index sensing[22–24] and optical modulators[4,6], etc.

For LN material, this method provides new methods to realize highly sensitive and ultracompact E-field sensors. In the past few years, LN nanostructures such as artificial metasurfaces and photonic crystals have been theoretically and experimentally investigated to enhance the sensitivity of E-field via quasi-BIC resonances[4,25–27]. However, LN is a material with stable physical and chemical properties, so it is difficult to accurately etch nanoscale structures[6,22]. Recently, nanostructures based on etchless LN has been confirmed to support quasi-BIC states and performed better in boosting nonlinear conversion efficiency[29,30]. Specifically, this way provides a platform for designing a highly sensitive E-field sensor operating at lower external voltage due to its strong optical confinement.

Here, we design and demonstrate a highly sensitive E-field sensor that can exhibit multiple functions in the near-infrared wavelength regime using etchless LN thin film via hybrid couple-grating system. Such devices are designed to support a quasi-BIC resonance with a quality (Q) factor of 2050, leading to strong localized electric field confinement inside LN layer. The efficient interaction between the optical field and external DC electric field causes the obvious change of LN refractive index. Therefore, our proposed structure achieved highly sensitive E-field sensor with a sensitivity of 40.8nm/V. And we estimate that the RF bandwidth of the E-field sensor should have broad RF bandwidth up to 150GHz by optimizing the interaction between optical confinement of quasi-BIC resonance and external electrodes. We believe that our proposed devices can be utilized in the field of extremely weak electric field signal detection.

## 2. Theory and Design of the proposed E-field sensor.

Fig.1 schematically illustrates the building blocks of our tunable hybrid coupling-grating nanostructures, consisting of an Au back-reflector, on top of which a lithium niobate (LN) layer is deposited. The LN layer acts as a dielectric spacer, adding a degree of freedom for the nanostructures. This layer is followed by deposition of an indium-tin-oxide (ITO) layer as external applied voltage control electrode and a DC electric bias is applied between the ITO layer and the Au layer[31]. On top of ITO layer, the SiO$_2$ nanograting serves as a bridge to promote the coupling between the waveguide mode of the LN layer and the incident light.

For LN material, its high nonlinear optical and EO properties is related to negative birefringence: ordinary refractive index $n_o$ and extraordinary refractive index $n_e$. So, these indices are affected by electric field. Consequently, its optical index variation with respect to external electric field $\vec{E} = (E_x, E_y, E_z)$ is linked to EO tensor as follows[10,32]:

$$\begin{vmatrix} \Delta\left(1/n^2\right) \\ \Delta\left(1/n^2\right) \\ \Delta\left(1/n^2\right) \\ \Delta\left(1/n^2\right) \\ \Delta\left(1/n^2\right) \\ \Delta\left(1/n^2\right) \end{vmatrix} = \begin{vmatrix} 0 & -r_{22} & r_{13} \\ 0 & r_{22} & r_{13} \\ 0 & 0 & r_{33} \\ 0 & r_{51} & 0 \\ r_{51} & 0 & 0 \\ -r_{22} & 0 & 0 \end{vmatrix} \begin{vmatrix} E_x \\ E_y \\ E_z \end{vmatrix} \quad (1)$$

Here, $r_{ij}$ are the elements of the EO tensor. For LN material, there are only four distinctive nonzero elements and $r_{33}$ ($r_{33}$=30.8 pm/V at λ=633 nm) is the largest element among them. To realize the obvious EO effect, one should utilize x-cut LN with crystalline $n_e$ along the y-axis while the $n_o$ in x- and z-axis in Cartesian right-angle coordinate system as shown in Fig.1. The refractive indexes of the semi-infinite SiO$_2$ substrate and gratings were 1.45[33]. The permittivity of Au layer and ITO layer were referred to Ref[33].and Ref[8]., respectively. As for the LN material,

its anisotropic and dispersive nature were described with the Sellmeier equation[34] as follows:

$$n_e = 1 + \frac{2.9804\lambda^2}{\lambda^2 - 0.02047} + \frac{0.5981\lambda^2}{\lambda^2 - 0.0666} + \frac{8.9543\lambda^2}{\lambda^2 - 416.08} \quad (2)$$

$$n_o = 1 + \frac{2.6734\lambda^2}{\lambda^2 - 0.01764} + \frac{1.2290\lambda^2}{\lambda^2 - 0.05914} + \frac{12.614\lambda^2}{\lambda^2 - 474.6} \quad (3)$$

To investigate the formation mechanism of our proposed E-field sensor, we preliminarily study the bandstructures for LN waveguide-gratings system utilizing finite-element methods (FEM, COMSOL Multiphysics) where the periodic boundary conditions are applied along x- and y- axis and perfect matched layer (PML) is applied along z-axis. The results of eigenfrequency analysis from visible to near-infrared range are shown in Fig.2(a-b), revealing that such a periodic system supports symmetry-protected BICs with an infinite Q factor at Γ point of the irreducible Brillouin zone. The degenerate eigenmodes at Γ point are decoupled from the external radiation continuum due to symmetry is broken, resulting from off-normal incidence[35–37]. Due to the slab geometry, the degenerated eigenmodes, i.e., quasi-BIC can be excited by TE-polarized light at off-normal incidence[37]. They perform the same in the bandstructures when compared with the one without 100nm-thick gold film, indicating that the gold film here operates as a total reflective mirror instead of modifying the optical modes supported by the devices (see Section 1-2 in Supplementary Materials). However, due to the loss of gold film, the BIC modes have transitioned into quasi-BIC resonance with the Q-factor becoming finite as shown in Fig.2(b, e).

Furthermore, we investigate the electric field distributions of quasi-BICs and plot them in Fig.2(c, f). It shows that the electric field is induced by the leaky guided mode and confined long the upper and lower surfaces of the entire megastructure. Particularly, the field distributions indicate that there are evanescent waves at the planar interface between the waveguide layer and substrates for the case with 100nm-thick gold film which provides strong light-matter interaction in electro-optical modulation. According to previous published works[38–40], such quasi-BIC resonance performances similarly with Fabry-Perot resonance which is very sensitive to the permittivity of materials, the length of cavity.

To further confirm the quasi-BIC states of our proposed structure, we utilize home-mode 3D-FDTD codes to calculate the reflective spectra with the normal incident TE-polarized plane wave. The results are shown in Fig.3(a), indicating that the quasi-BIC resonance with low quality factor is excited while the target high-Q resonances discriminate from the normal-incident reflectance spectra due to the presence of the Fabry-Pérot interference between LN waveguide and SiO$_2$ nanograting[35,41–43]. To evaluate the Q factor of the quasi-BIC resonance, we do a Lorentzian formula fitting[44–46] in Fig.3(b) and (c) for quasi-BIC2 and quasi-BIC1 with Q factors of 220 and 2050, respectively.

## 3. Principle of Electric Field Sensing based on Thin Film Lithium Niobate.

To investigate the largest EO effect, one shall overlap the largest E-field component with $r_{33}$ which along the crystalline of LN. Thus, making the tensor reflecting anisotropic characteristics. By neglecting other small EO element $r_{31}$, $r_{51}$ and $r_{22}$, the EO effect of LN can be described as[47,48]

$$\Delta n_e = -\frac{1}{2}n_e^3 r_{33}\frac{U}{d} \quad (4)$$

Where $n_e$ is the extraordinary refractive index of LN, which is 2.17 at wavelength of 950 nm. The $U$ represents the external voltage between the gold layer and ITO layer and $d$ is the thickness of LN layer in this work. However, the distance between the two electrodes is at the nanometer level, which can generate great electric field strength. For this resonantly enhanced device in which light slowly decays, nonlinear effects could be realized significantly. In our proposed device, we focus on the resonantly enhancement of electro-optic effect of LN layer. In Ref[11,25,49,50], the optical field localization factor of an EO modulator or E-field sensor based on LN nanostructures can be estimated with the ensemble field enhancement which is denoted as $\overline{f_{opt}}$

$$\overline{f_{opt}} = \frac{\int_{patterned} \left| E\left(x,y,z\right) \right| dV}{\int_{unpatterned} \left| E\left(x,y,z\right) \right| dV} \qquad (5)$$

Here, the numerator represents the electric field amplitude integration over LN materials region in the patterned nanostructures while the denominator is the same quantity obtained for unpatterned LN thin film. To characterize the optical field factor more accurately, we utilize three-dimensional simulation to calculate the optical field integration of our proposed configuration with/without the hybrid coupling-grating system. The results are shown in Fi.4, and we calculated the $\overline{f_{opt}}$ up to 14.2 and 5.4 for quasi-BIC1and quasi-BIC2, respectively, demonstrating that the proposed structure with hybrid coupling-grating system. In order to theoretically analyze the $\Delta\lambda_{res}$ sensitivity with respect to the measured, $\overline{f_{opt}}^2$ is integrated in Eq.2 (like the $\overline{f_{opt}}$) in order to quantify the light enhancement induced $r_{33}^{eff}$. Consequently, the corresponding induce refractive index local variation in PhC structure can be expressed as:

$$\Delta n_e\left(x,y,z\right) = -\frac{1}{2} n_e^3 f_{opt}^2\left(x,y,z\right) r_{33} \frac{U}{d}$$

This sensitivity analysis employing $\overline{f_{opt}}$ method had been utilized in sensitivity analysis of SPL guided resonance-based temperature sensor and well agreement between simulations and experiments was achieved. Therefore, we will employ here the same method for accurate sensitivity of our proposed E-field sensor.

To get a clear idea of the influence of the dielectric's refractive index variation on the quasi-BIC resonance above, we next investigate the optical spectra plotted in Fig.4(a) by changing $n_e$ from 2.11 to 2.23 with a step of 0.02. The results indicate a shift of filtering wavelength toward the longer wavelength regime. Noting that the relation between applied electric field strength and the applied voltage $U = Ed$ , where $d$ is the distance between the anode and the cathode, we can see that the refractive index and thus the effective thickness of the LN layer increase as the applied voltage increases, which means that the resonance wavelength of the Fabry-Perot resonance increases with the applied voltage, giving an explanation to the results obtained.

To calculate the applied voltage corresponding to refractive index variation $\Delta n_e$, we take $\overline{f_{opt}}$, $d=300$nm, $n_e=2.17$ and $r_{33}=30.8$pm/V into account and substitute into Eq.3 to get the calculation results as $\Delta n_e = 0.1058U$ and $\Delta n_e = 0.0153U$ for quasi-BIC1 and quasi-BIC2. respectively. Fig.4(b-c) shows the relationship between the F-P resonant wavelength and reflection ratio and the applied voltages. For quasi-BIC1, the refractive index $n_e$ is changed from 2.11 to 2.23, corresponding the applied voltage from -0.57 to +0.57V while for quasi-BIC2, the applied voltage varies from -3.93 to +3.93V. The filtering wavelength linearly shifts from 918 to 980nm. The reason to this result is that the resonant wavelength supported by the FP cavity is closely related to the length of the cavity and the refractive index of the material in the cavity[23,35].

The basic parameters of the E-field sensor are tuning sensitivity (S) and FOM[6,51], which are, respectively, calculated from

$$S = \Delta\lambda / \Delta V \qquad (6)$$

$$FOM = S/FWHM \qquad (7)$$

Where $\Delta\lambda$ is the resonance wavelength change, $\Delta V$ is voltage change, and FWHM is the full width at the half maximum of spectrum. As shown in Table1, the quasi-BIC1 has a tuning sensitivity of 40.8nm/V which is about 12 times larger than that of quasi-BIC2. And the FOM of quasi-BIC1 is also 80 times larger than that of quasi-BIC2.The reason to these results is that quasi-BIC1 performs strong optical field confinement, leading to efficient interaction between optical field and applied electric field.

**Table 1 The basic parameters of the proposed E-field sensor**

| Mode Profiles | Tuning sensitivity (S) | Figure of merits (FOM) |
|---|---|---|
| Quasi-BIC 1 | 40.8 nm/V | 88.05 |
| Quasi-BIC 2 | 3.53nm/V | 1.06 |

To calculate the minimum detectable E-field of the sensor, we take the resolution of commercially available OSA spectroscopy into consideration and it is 1nm. Therefore, we can obtain the minimum E-field of about 1.04V, corresponding to an electric field strength of $3.47 \times 10^7$V/m applied on the longitudinal direction of LN layer. Consequently, our proposed E-field sensor device based on etchless LN via coupling-grating system is estimated able to detect a minimum E-field as small as $3.47 \times 10^7$V/m. According to the parameters comparison in Table 1, our proposed E-field sensor shows good performance in extremely weak electric field detection.

**Table 2 Parameters comparison of E-field sensors based on LN material**

| Structure | E-field sensitivity | E-field amplitude for wavelength shift 1nm |
|---|---|---|
| LN Photonic crystal[32] | 0.19 pm/V | 5270V |
| APE LN photonic crystal[11] | 0.6 nm/V | 1.67V |
| Periodical Au nanoparticle on LN layer[51] | 0.25 nm/V | 4V |
| Tunable LN metasurfaces[26] | 0.07nm/V | 14.3V |
| 1D metasurfaces with LN nanograting[52] | 0.72 nm/V | 1.4V |
| In this work | 40.8 nm/V (at 953nm) | 24.5 mV |
| | 3.53 nm/V (at 733nm) | 283.3 mV |

## 4. The Estimation of the Modulation depth and the RF bandwidth.

To calculate the modulation depth of our proposed E-field sensor, we simulate the reflection spectra when the refractive index variations is $\Delta n_e=0.002$, corresponding to the applied voltages of $U=18.91$mV. The result is shown in Figure5, indicating that the resonant wavelength shifts 0.38nm, and the modulation depth $\eta$ reaches to 80% under different biases. We quantify the performance of our E-field sensor by their modulation strength[53,54] $\eta$

$$\eta = \frac{R(V) - R(0)}{R(0)} \qquad (8)$$

Where the reflectance difference of $R(0)$ and $R(V)$, where $R(V)$ is reflectance at the driving voltage $V=18.91\text{mV}$, and $R(0)$ is the reflectance at 0V.

In a resonator, the modulation bandwidth is limited by the permittivity of materials, photon lifetime, walk-off between electric and optical signals and RF (radio frequency) electrodes. Therefore, we estimate the 3dB modulation bandwidth ($f_{3dB}$) with the RC (Resistor-Capacitance) time and the photon lifetime as followed:

$$f_{3dB} = \sqrt{\frac{1}{\left(\frac{\lambda Q}{c_0}\right)^2 + \left(2\pi RC\right)^2}} \qquad (9)$$

Here, $\lambda$ and $Q$ are the wavelength and the quality factor of resonance; $c_0$ is the light speed in vacuum. R and C are the contact resistance and the capacitance of the device. As a result, the $f_{3dB}$ response can be estimated with photon lifetime required to build up and release the energy from the resonator. From the obtained Q-factor of 2050, the $f_{3dB}$ was calculated to be approximately 154GHz while that of quasi-BIC2 is 1.86THz. In previous works, the theoretical frequency responses of sandwich-like nanostructured optical modulator are determined by the lifetime of photons in cavity[6,55]. However, the defects of the electrode and the loss in the electrical connections are the mainly reasons that causes the bandwidth reduction in experiment.

## 5. Conclusion

In summary, we demonstrated that a highly sensitive E-field sensor based on the etchless LN material can generate quasi-BIC resonance with a high-quality factor of 2050 via the hybrid coupling-grating system. Such a high-Q resonance provides strong localized optical field confinement inside LN material which strengthens the interaction between applied electric field and optical field and a normalized optical field factor is obtained. Based on the analysis above, our proposed E-field sensor with tuning sensitivity of 40.8nm/V and minimum detectable electric field amplitude of 24.5mV with the resonant wavelength shift of 1nm has been realized. By better designing the electrode, the RF bandwidth can reach 154GHz. Based on the analysis above, we believe that such an E-field sensor will be employed in the field of extremely weak electric field detection.

## Supplementary Materials

The supporting materials are published as submitted, without typesetting or editing. The responsibility for scientific accuracy and content remains entirely with the authors

## Declaration of competing interest

The authors declare that they have no known competing financial interests or personal relationships that could have appeared to influence the work reported in this paper

## Funding sources

We obtain the financial supports from the High-level Talents Project of Guangdong Mechanical & Electrical Polytechnic (Gccrcxm-202208).

## Acknowledgement

We also thank Pro. Huihui Lu from Jinan University, Guangzhou China for providing the usage of home-mode FDTD codes and COMSOL Multiphysics software in this paper.

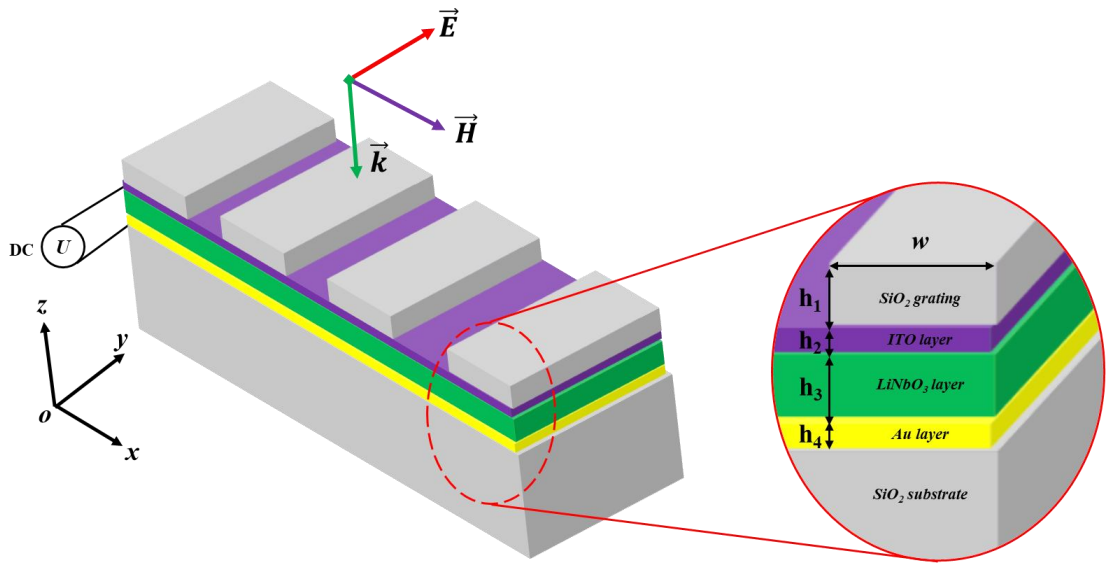

Fig.1. Unit cell design of our proposed E-field sensor. Schematic of (a) periodic array and (b) unit cell of the antenna elements. The E-field sensor is composed of a SiO₂ coupling grating, an ITO layer, a LN layer, and an Au back-reflector on top of semi-infinite SiO₂ substrate. The periodicity of coupling grating is Λ=510 nm, and the thickness of the Au back reflector, LN, ITO layers are $h_2$=100nm, $h_3$=300nm, $h_4$=100nm, respectively. The width and thickness of the SiO₂ gratings are $w$=300nm, and $h_1$=200nm, respectively.

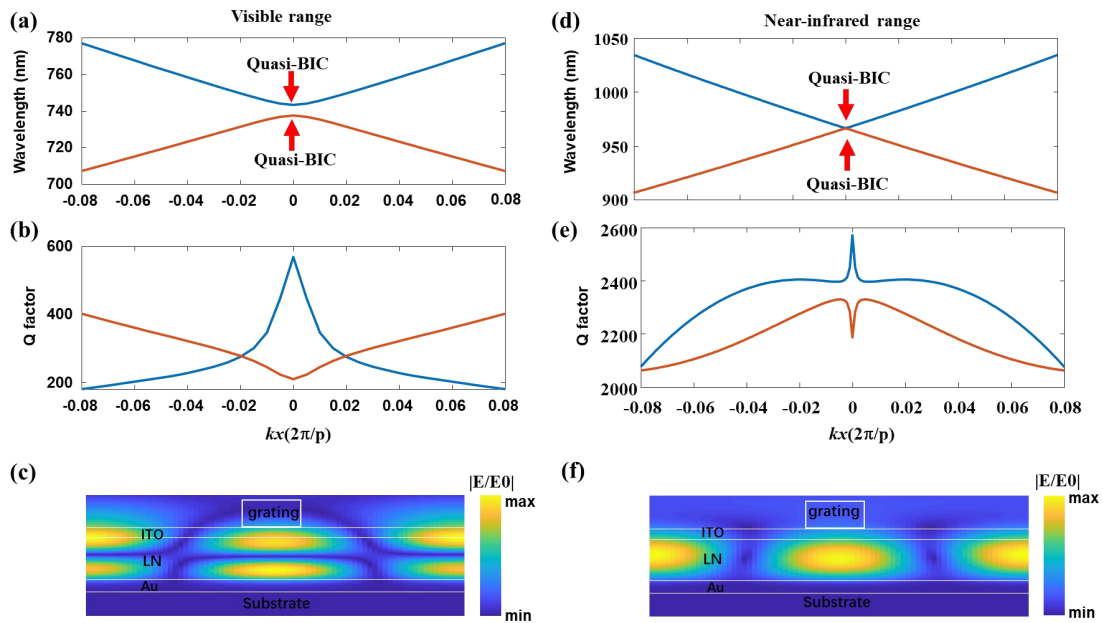

Fig.2. (a, d) Band structure of the E-field sensor and (b, e) the quality factor. (c, f) the corresponding electromagnetic field distribution of eigenmodes at Γ point. The structural parameters are: Λ=510nm, $h_1$=200nm, $h_2$=100nm, $h_3$=400nm, $h_4$=100nm and $w$=300nm.

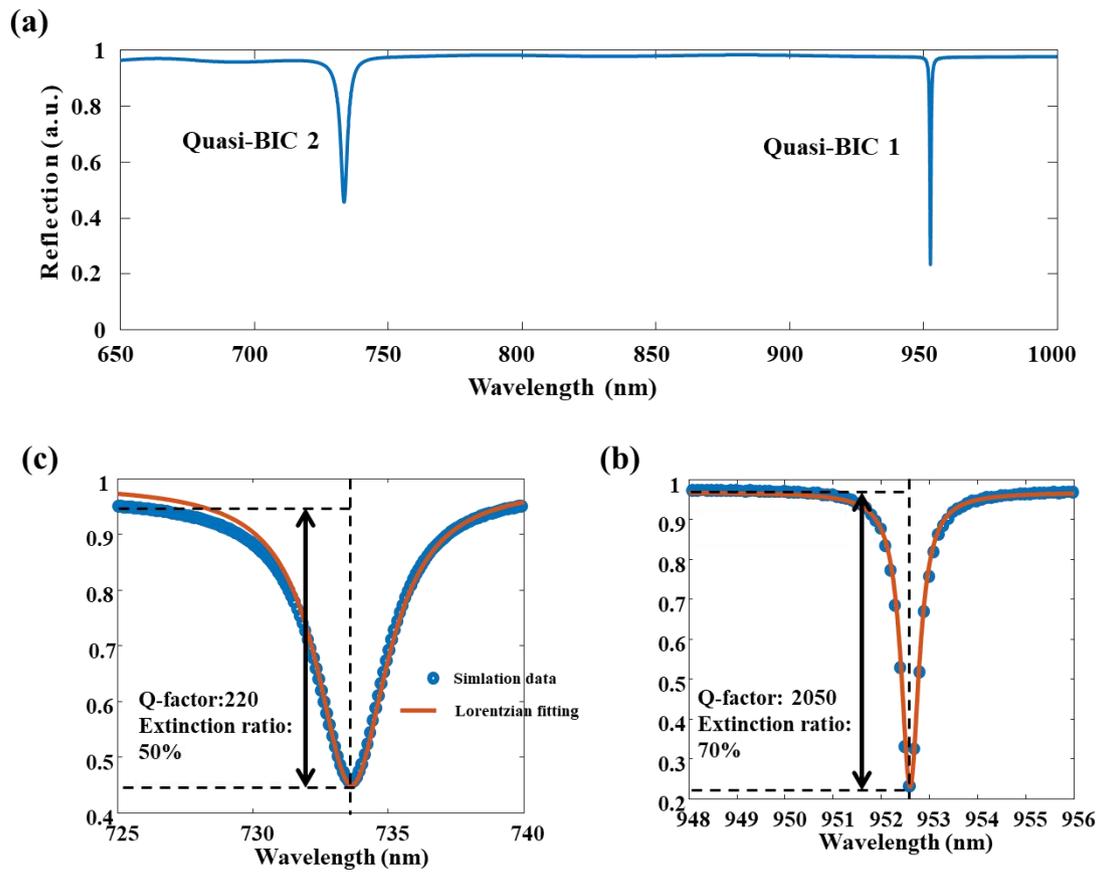

Fig.3. (a) The reflective spectra of the E-field sensor from visible to near-infrared range with the normal TE-polarized incident light. (b-c) the Lorentzian lineshape fitting of quasi-BIC1 and quais-BIC2.

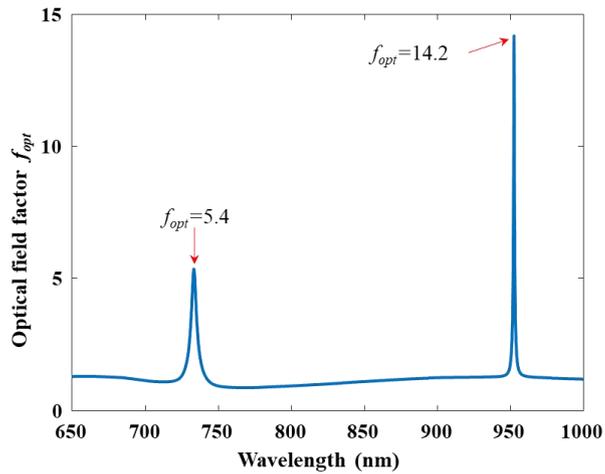

Fig.4. The optical field factor of the E-field sensor.

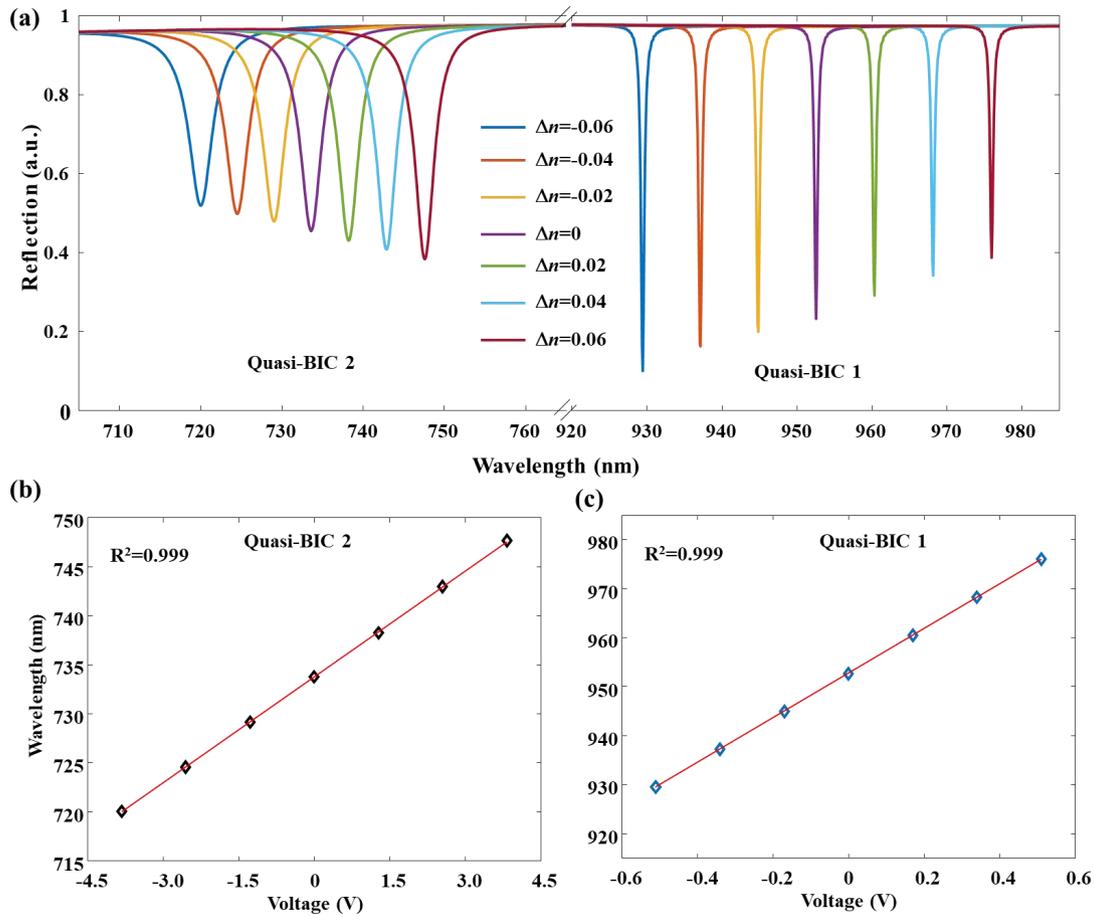

Fig.4 (a) Optical spectra produced by the variation of $\Delta n_e$ from -0.06 to 0.06. (b-c) Resonant wavelength shifts of quasi-BIC1 and quasi-BIC2. The applied voltages for quasi-BIC2 are from -3.95 to 3.95V while that of quasi-BIC1 are -0.57 to 0.57V.

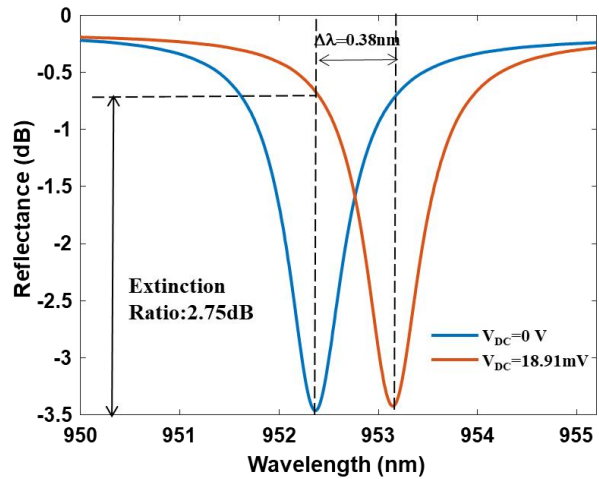

Fig.5 The modulation principle of the proposed E-field sensor. The wavelength shift is 0.38nm when the applied voltage is 18.91mV and modulation depth reach 80% when the applied voltage tuned from 0 to 18.91mV.